\documentclass[aps,prb,twocolumn,showpacs,preprintnumbers,amsmath,amssymb,superscriptaddress]{revtex4-1}
\usepackage{color}
\usepackage{graphicx}
\usepackage{dcolumn}
\usepackage{bm}

\usepackage[hypertex]{hyperref}
\newcommand{\nc}{\newcommand}
\nc{\be}{\begin{eqnarray}}
\nc{\ee}{\end{eqnarray}}
\nc{\bea}{\begin{eqnarray}}
\nc{\eea}{\end{eqnarray}}
\nc{\bean}{\begin{eqnarray*}}
\nc{\eean}{\end{eqnarray*}}
\nc{\mb}{\mbox}
\nc{\rnc}{\renewcommand}
\nc{\vk}{\mb{\boldmath$k$}}
\nc{\vx}{\mb{\bf x}}
\nc{\br}{\mb{\bf r}}
\nc{\bv}{\mb{\bf v}}
\nc{\bp}{\mb{\bf p}}
\nc{\ve}{\mb{\bf e}}
\nc{\vz}{\hat {\mb{\bf z}}}
\nc{\vp}{\mb{\boldmath$p$}}
\nc{\vb}{\mb{\boldmath$b$}}
\nc{\rr}{\mb{\boldmath$r$}}
\nc{\vR}{\mb{\boldmath$R$}}
\nc{\vj}{\mb{\boldmath$j$}}
\nc{\vg}{\mb{\boldmath$g$}}
\nc{\vm}{\mb{\boldmath$m$}}
\nc{\vd}{\mb{\boldmath$d$}}
\nc{\hd}{\mb{\boldmath$\hat{d}$}}
\nc{\vD}{\mb{\boldmath$D$}}
\nc{\vF}{\mb{\boldmath$F$}}
\nc{\vG}{\mb{\boldmath$G$}}
\nc{\vI}{\mb{\boldmath$I$}}
\nc{\vW}{\mb{\boldmath$W$}}
\nc{\x}{\mb{\boldmath$x$}}
\nc{\A}{\mb{\boldmath$A$}}
\nc{\va}{\mb{\boldmath$a$}}
\nc{\vv}{\mb{\boldmath$v$}}
\nc{\vq}{\mb{\boldmath$q$}}
\nc{\vn}{\mb{\boldmath$n$}}
\nc{\vJ}{\mb{\boldmath$J$}}
\nc{\vS}{\mb{\boldmath$S$}}
\nc{\vs}{\mb{\boldmath$\sigma$}}
\nc{\vE}{\mb{\boldmath$E$}}
\nc{\vB}{\mb{\boldmath$B$}}
\nc{\vM}{\mb{\boldmath$M$}}
\nc{\vL}{\mb{\boldmath$L$}}
\nc{\vpsi}{\mb{\boldmath$\psi$}}
\nc{\vphi}{\mb{\boldmath$\varphi$}}
\nc{\Vphi}{\mb{\boldmath$\phi$}}
\nc{\Vomega}{\mb{\boldmath$\Omega$}}
\nc{\ipsi}{\it{\Psi}}
\nc{\vepsilon}{\mb{\boldmath$\epsilon$}}
\nc{\valpha}{\mb{\boldmath$\alpha$}}
\nc{\vgamma}{\mb{\boldmath$\gamma$}}
\nc{\vomega}{\mb{\boldmath$\omega$}}
\nc{\vmu}{\mb{\boldmath$\mu$}}
\nc{\vt}{\mb{\boldmath$\tau$}}
\nc{\vT}{\mb{\boldmath$T$}}
\nc{\vpi}{\mb{\boldmath$\pi$}}
\nc{\nab}{\nabla}
\nc{\ov}{\overline}
\nc{\cdott}{\!\cdot\!}
\nc{\cdottt}{\!\!\cdot\!}
\nc{\LL}{\Big{\langle}}
\nc{\RR}{\Big{\rangle}}
\nc{\LR}{\Bigm{|}\!}
\nc{\vP}{\mb{\boldmath$P$}}
\nc{\nnn}{\nonumber\\}

\begin{document}

\title{
Quantum thermal Hall effect of Majorana fermions\\ on the surface of 
superconducting topological insulators
}

\author{Yosuke Shimizu}
\affiliation{
Institute for Materials Research, Tohoku University, Sendai 980-8577, Japan
}
\author{Ai Yamakage}
\affiliation{
Department of Applied Physics, Nagoya University, Nagoya 464-8603, Japan
}
\author{Kentaro Nomura}
\email{nomura@imr.tohoku.ac.jp}
\affiliation{
Institute for Materials Research, Tohoku University, Sendai 980-8577, Japan
}
\date{\today}

\begin{abstract}
We study the quantum anomalous thermal Hall effect in a topological superconductor
 which possesses an integer bulk topological number, and supports Majorana excitations on the surface. 
To realize the quantum thermal Hall effect, a finite gap at the surface is induced by applying an external magnetic field or by the proximity effects with magnetic materials or $s$-wave superconductors with complex pair-potentials. 
Basing on the lattice model Hamiltonian for superconducting states in Cu-doped Bi$_2$Se$_3$, we compute the thermal Hall conductivity as a function of various parameters such as the chemical potential, the pair-potential, and the spin-orbit coupling induced band gap.
It is argued that the bulk topological invariant corresponds to the quantization rule of the thermal Hall conductivity induced by complex $s$-wave pair-potentials.
\end{abstract}

\maketitle


\section{Introduction}
Topological insulators and superconductors are new quantum states of matter, characterized by topological numbers\cite{review_TI}.
The quantum Hall effect (QHE)\cite{review_QHE} is a first found topologically nontrivial state, where the Hall conductivity is quantized as
\bea
 \sigma_{xy}=\nu \frac{e^2}{h},
\eea
$\nu$ being an integer value corresponding to 
the topological number of bulk wave functions
\cite{TKNN}.  
The two-dimensional 
(2d) topological superconductors and superfluids
with chiral ($p$-wave) Cooper pairing are
superconductor analogues of the QHE and considered to be realized,
e.g.
in a thin film of $^3$He A phase
\cite{Leggetbook,Volovik}, Sr$_2$RuO$_4$
\cite{Maeno2003}, and the $5/2$-filling
fractional QHE
\cite{Moore-Read}.
In superconductors\cite{TSF}, charges are not conserved and thus electric transport study such as quantum Hall measurement cannot characterize their topological nature.
Instead, since the energy is still conserved, thermal transport especially the thermal Hall conductivity, reflects the topological character of topological superconductors as\cite{Read2000}
\bea
 \kappa_{xy}=\nu\frac{\pi^2k_B^2}{6h}T.
\eea
Here $\nu$ corresponds to the bulk topological number.\cite{Read2000}

Recent 
studies have 
shown that topological 
states exist in time-reversal invariant 
and 
three-dimensional
(3d)
cases
as well \cite{review_TI}.
The surface of three-dimensional topological insulators supports gapless excitations with a linear dispersion, for some simple cases, which can be described by the two-dimensional massless Dirac Hamiltonian.
One of the intriguing phenomena is the quantum anomalous Hall effect on the surface.\cite{Fu2007,Qi2008,Yu2010,Nomura2011,Zhou2013}
Recently, the experimental observation of the quantum anomalous Hall effect was reported in Cr-doped
(Bi,Sb)$_2$Te$_3$ thin films\cite{Exp_QAHE}, where the Hall conductivity exhibits a clear quantized plateau, accompanied by a considerable drop in the longitudinal resistance.

Theoretically the systematic
classification of topologically nontrivial insulators and superconductors is 
established in terms of symmetries and dimensionality,
and clarified that topologically nontrivial superconductors (TSCs) and superfluids (TSFs) with time-reversal symmetry are also realized in three-dimensions\cite{Schnyder2008,Kitaev2009}.
In contrast to three-dimensional topological insulators which are characterized by ${\mathbb Z}_2$ topological numbers, three-dimensional topological superconductors are characterized by integers ${\mathbb Z}$.\cite{Schnyder2008,Kitaev2009}
An example of 3d-TSFs is the B phase
of superfluid $^3$He \cite{Schnyder2008}. 
From the bulk-boundary correspondence, there exist
topologically protected gapless Andreev bound states in TSCs. In particular, the superconductivity
infers that the gapless Andreev bound states are their own antiparticles, thus Majorana fermions.\cite{Schnyder2008}
On the surface of $^3$He B phase, a clear linear dispersion of in-gap states was measured experimentally\cite{Murakawa2011}.
In addition, the newly found superconducting phase in Cu-doped Bi$_2$Se$_3$
\cite{Hor2010} 
has been proposed to be a 3d-TSC
\cite{Fu2010}.
Recently, point-contact spectroscopy experiments in Cu-doped Bi$_2$Se$_3$
\cite{Sasaki2011,Kirzhner2012}
 have reported a zero-bias conduction peak which is in debate whether it indicates gapless surface modes with characteristic dispersion relations modified from the linear Majorana cone\cite{Hsieh2012,Yamakage2012} or not\cite{Peng2013}.

As surface Dirac fermions realize the anomalous quantum Hall effect in topological insulators,
the anomalous quantum thermal Hall effect of Majorana fermions occurs on the surface of topological superconductors.
For the ideal case where surface spectra have a linear Dirac-Majorana dispersion such as $^3$He B phase, the thermal Hall conductivity can be easily evaluated\cite{Ryu2011,Wang2011,Nomura2012,Shiozaki2014} as
$
  \kappa_{xy}=\pm \frac{\pi^2k_B^2}{12h}T.
$
On the other hand, it is not obvious what the value of the thermal Hall conductivity is in the case of Cu-doped Bi$_2$Se$_3$ with modified surface dispersions.
In addition, it would be natural to question whether the quantized thermal Hall conductivity is related to the bulk topological number.

In this work we study anomalous thermal transport on the surface of superconducting doped topological insulators basing on a microscopic model for Cu$_x$Bi$_2$Se$_3$.\cite{Fu2010} We first calculate the phase diagram of the effective Bogoliubov-de Genne Hamiltonian for the bulk case, as a function of the original band gap of the host material, the chemical potential, and the superconducting pair-potential. There exist different types of gapped superconducting states. 
We characterize the gapped states by computing the topological numbers in the bulk and the thermal Hall conductivity in a slab geometry.

The paper is organized as follows: In section II we present the model Hamiltonian for bulk superconducting states of Cu-doped Bi$_2$Se$_3$.
We study the ground state phase diagram in section III, and the bulk topological numbers in section IV as a function of the chemical potential, the pair-potential, and the band gap.
In section V the model Hamiltonian for a slab geometry is introduced and the surface energy spectra are calculated.
In section VI the anomalous thermal Hall conductivity is computed and compared with the bulk topological number.
In section VII we present a summary of the results obtained in this work.

\section{Model Hamiltonian}
In the presence of translational invariance, the superconducting state of Cu$_x$Bi$_2$Se$_3$ is described by the momentum-space Bogoliubov-de Genne Hamiltonian\cite{Fu2010,Kirzhner2012,Sasaki2011}
\bea
 H_{\rm BdG}=\sum_{\vk}[c^{\dag}_{\vk},c^{T}_{-\vk}]\,{\cal H}_{\rm BdG}(\vk)
\begin{bmatrix}
c^{}_{\vk} \\ c^{\dag T}_{-\vk}
\end{bmatrix},
\label{hamiltonian-1}
\eea
where $c^{\dag}_{\vk}=(c^{\dag}_{\vk\uparrow -},c^{\dag}_{\vk\downarrow -},c^{\dag}_{\vk\uparrow +},c^{\dag}_{\vk\downarrow +})$ and
$c^{T}_{\vk}=(c^{}_{\vk\uparrow -},c^{}_{\vk\downarrow -},c^{}_{\vk\uparrow +},c^{}_{\vk\downarrow +})$
are the creation and annihilation operator for electrons, $\pm$ is the band index showing parity, and
\bea
 {\cal H}_{\rm BdG}(\vk) &=&
\begin{bmatrix}
h(\vk) & \hat\Delta \\
\hat\Delta^{\dag} & -h^T(-\vk)
\end{bmatrix}
\label{hamiltonian-2}
\eea
is an $8\times8$ matrix.
The diagonal part $h(\vk)$ describes the normal electronic states of Cu-doped Bi$_2$Se$_3$.
For simplicity we consider the highest valence band and lowest conduction band\cite{BiSe-model}, each band has two-fold degeneracy in the presence of time-reversal and inversion symmetries.
As a simplified model, we consider the hexagonal lattice with the lattice constants $a$ and $c$, where the 2D triangular lattices stack along the c-axis direction.
Although the real crystal structure is more complicated, this model describes low-energy physics of the system.
The effect of doping Cu is taken account as the shift of the chemical potential.
The 4 band Hamiltonian $h(\vk)$ is given by
\bea
 h(\vk)=\sum_{i=1}^3R_i(\vk)\alpha_i+M(\vk)\beta+\epsilon(\vk)I
\eea
where
\bea
 \alpha_i=
\begin{pmatrix}
0 & \sigma_i \\ \sigma_i & 0
\end{pmatrix},
\quad
 \beta=
\begin{pmatrix}
I & 0 \\ 0 & -I
\end{pmatrix},
\eea
are $4\times4$ Dirac matrices,
$\sigma_i$ being the Pauli matrix, $I$ identity matrix,
and\cite{BiSe-model,Fu2010,Sasaki2011,Hsieh2012,Yamakage2012,Hashimoto2013,Kirzhner2012}
\bea
R_1(\vk)&=&A_1\frac{2}{\sqrt{3}}\sin\Big(\frac{\sqrt{3}}{2}k_xa\Big)\cos\Big(\frac{1}{2}k_ya\Big) \nnn
R_2(\vk)&=&A_1\frac{2}{3}\Big[\cos\Big(\frac{\sqrt{3}}{2}k_xa\Big)\sin\Big(\frac{1}{2}k_ya\Big)+\sin\Big(k_ya\Big)\Big] \nnn
R_3(\vk)&=&A_3\sin\Big(k_zc\Big) \nnn
M(\vk)&=&M_0-B_1\Big[2-2\cos\Big(k_zc\Big)\Big]-\frac{4}{3}B_2\Big[3-\cos\Big(k_ya\Big) \nnn
&& \qquad\qquad\qquad
-2\cos\Big(\frac{\sqrt{3}}{2}k_xa\Big)\cos\Big(\frac{1}{2}k_ya\Big)\Big] \nnn
\epsilon(\vk)&=&
-\mu+D_1\Big[2-2\cos\Big(k_zc\Big)\Big]+\frac{4}{3}D_2\Big[3-\cos\Big(k_ya\Big) \nnn
&& \qquad\qquad\qquad 
-2\cos\Big(\frac{\sqrt{3}}{2}k_xa\Big)\cos\Big(\frac{1}{2}k_ya\Big)\Big].
\eea
The band gap is given by the parameter $M_0$.
In this notation, $M_0>0\ (<0)$ corresponds to a topological (ordinary) insulator. 
For Bi$_2$Se$_3$ we use the following values:
$A_1=1.007$[eV], 
$A_3=0.32$[eV], 
$B_1=0.216$[eV], 
$B_2=3.41$[eV], 
$D_1=0.024$[eV], 
$D_2=1.18$[eV]\cite{BiSe-model,Sasaki2011}, 
The chemical potential is measured from the Dirac point.

For the pair-potentials, it has been argued that four types of pair potentials could be realized within the irreducible representation for the $D_{3d}$ point group, $\hat\Delta_1$, $\hat\Delta_2$, $\hat\Delta_3$, and $\hat\Delta_4$.\cite{Fu2010,Kirzhner2012}
In the following, among these four types, we focus on the topologically nontrivial one with a full gap, $\hat\Delta_2$, which can be express as
\bea
 \hat\Delta_2=
\Delta
\begin{pmatrix}
0 & i\sigma_y \\
i\sigma_y & 0
\end{pmatrix}.
\eea
In this work, we do not solve the self-consistent equation, but introduce $\Delta$ as a tuning parameter.

\begin{figure}[t]
\begin{center}
\includegraphics[width=10cm]{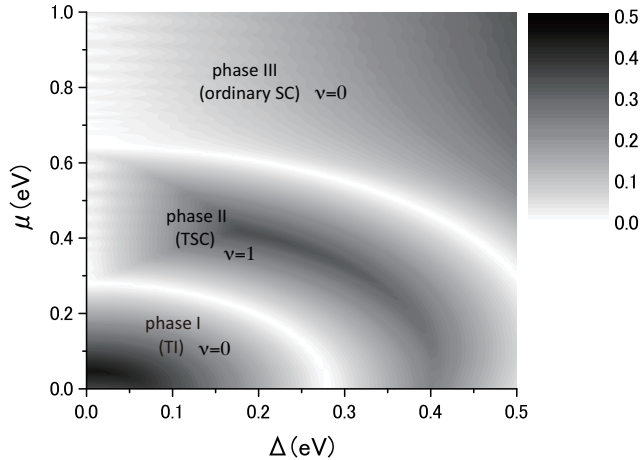}
\caption{Excitation gap as a function of the chemical potential $\mu$ and the pair-potential $\Delta$, where the band gap fixed at $M_0=0.28$ [eV].
}
\label{mu-delta}
\end{center}
\end{figure}

\begin{figure}[t]
\begin{center}
\includegraphics[width=10cm]{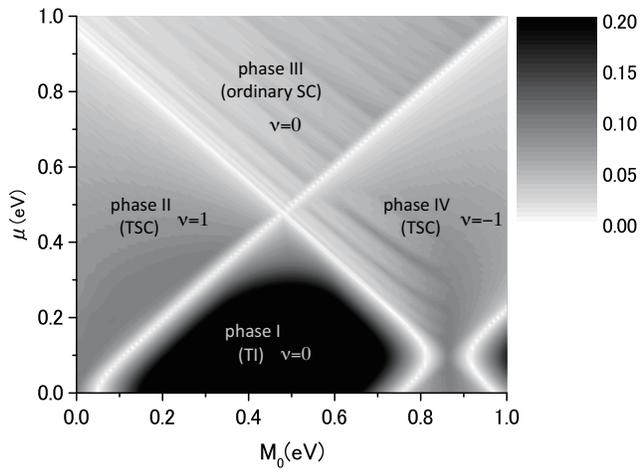}
\caption{Excitation gap as a function of the chemical potential $\mu$ and the band gap parameter $M_0$. The pair-potential is fixed at $\Delta=0.05$ [eV].
}
\label{mu-m0}
\end{center}
\end{figure}

\section{Phase diagrams}

We first study the phase structures of the BdG Hamiltonian eq.(\ref{hamiltonian-2}).  
Figure \ref{mu-delta} shows the excitation gap of the bulk spectrum of the Hamiltonian eq.(\ref{hamiltonian-2}) as a function of the chemical potential $\mu$ and the superconducting pair-potential $\Delta$. The band gap parameter is fixed at $M_0=0.28$ [eV].
The phase transitions are signaled by the vanishing excitation gap in Fig. \ref{mu-delta}.
In the normal case, $\Delta=0$, the excitation gap is finite only when the chemical potential is in the original band gap, $|\mu|\le M_0$ (phase I).
When the chemical potential is above the band gap, $|\mu| > M_0$, the system is in the metallic phase. The finite excitation gap is induced by the nonzero pair-potential $\Delta$.
Within the parameter region of Fig. \ref{mu-delta}, there are two superconducting phases, phase II and phase III, distinguished by gap vanishing. In the following, we will see that one of the superconducting state (phase II) is topologically nontrivial while the other (phase III) is trivial. 
In phase I, since the chemical potential $\mu$ is located in the original band gap, and thus the density of states vanishes, it is hardly expected that the superconducting order is developed spontaneously.
In this sense, phase I is an artifact introduced by assuming a finite bulk pair-potential.
Nevertheless it is of theoretical interest to see how two Majorana dispersions contribute to the thermal Hall conductivity as discussed in VI.

Figure \ref{mu-m0} shows the excitation gap as a function of the chemical potential $\mu$ and the band gap $M_0$, where the pair-potential is fixed at $\Delta=0.05$ [eV]. 
This result indicates that beside the topological insulating state (phase I), there are three superconducting gapped states (phase II, III, and IV) which are separated by gap closing.

\begin{figure}[b]
\begin{center}
\includegraphics[width=0.3\textwidth]{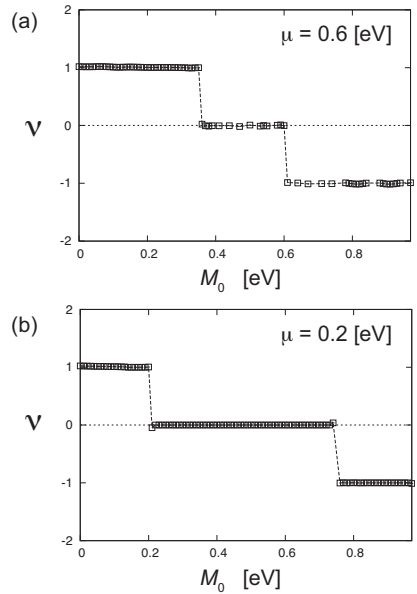}
\caption{
Winding number defined by eq. (\ref{winding-number-def}) is plotted as a function of the original band gap $M_0$.
The chemical potential is fixed at (a) $\nu=0.6$ [eV] (a) and (b) $\nu=0.2$ [eV].
}
\label{nu}
\end{center}
\end{figure}

\section{Symmetries and topological numbers}

Here we study the topological properties of the phases appeared in the phase diagram by computing the topological numbers.
The Bogoliubov-de Genne Hamiltonian has both time-reversal ($\Theta$) and particle-hole ($\Xi$) symmetries,
\bea
 \Theta^{-1}{\cal H}_{\rm BdG}(\vk)\Theta\ &=&\ \ {\cal H}_{\rm BdG}(-\vk)
\\
 \Xi^{-1}{\cal H}_{\rm BdG}(\vk)\Xi\ &=&-{\cal H}_{\rm BdG}(-\vk).
\eea
Because of these symmetries, one can find a unitary matrix $\Gamma=\Theta\Xi$ that anticommutes with the Hamiltonian\cite{Schnyder2008}
\bea
 \Gamma^{-1}{\cal H}_{\rm BdG}(\vk)\Gamma &=&-{\cal H}_{\rm BdG}(\vk).
\label{chiral-symmetry}
\eea
This chiral symmetry implies that eigenenergies of the Bogoliubov-de Genne Hamiltonian (eq. (\ref{hamiltonian-2}))
appear as pairs of $\pm E_n(\vk)$:
\bea
 {\cal H}_{\rm BdG}(\vk)\LR u^{\pm}_n(\vk)\RR
&=&
\pm E_n(\vk) \LR u^{\pm}_n(\vk)\RR
\eea
Here we introduce the $Q$-matrix defined as
\bea
 Q(\vk)& =& \sum_n\Big( 
\LR u^{+}_n(\vk)\RR \LL u^{+}_n(\vk)\LR
\nnn
&&\qquad\qquad
-
\LR u^{-}_n(\vk)\RR \LL u^{-}_n(\vk)\LR
\Big).
\eea
As a consequence of chiral symmetry eq.(\ref{chiral-symmetry}), the $Q$ matrix can be brought into block off-diagonal form,
\bea
 Q(\vk)&=&
\begin{bmatrix}
0 & q(\vk) \\
q^{\dag}(\vk) & 0
\end{bmatrix}, \quad q(\vk)\in U(4)
\eea
  in the basis in which $\Gamma$ is diagonal.
The topological index in the presence of chiral symmetry is given by the winding number\cite{Schnyder2008}
\bea
 \nu
&=&
\int \frac{d^3k}{24\pi^2}\epsilon_{\mu\nu\rho}
{\rm tr}\Big[
(q^{-1}\partial_{\mu}q)(q^{-1}\partial_{\nu}q)(q^{-1}\partial_{\rho}q)
\Big]
\label{winding-number-def}
\eea
which takes integer values,
where $\mu, \nu, \rho=k_x, k_y, k_z$, and the integral extends over the entire Brillouin zone.
As shown in Fig. \ref{nu} the evaluated values of $\nu$ are 0, 1, 0, and $-1$ in phase I, II, III, and IV, respectively.

\section{Surface modes of topological superconductors}

To study the energy dispersion of surface modes, and compute the thermal Hall conductivity,
we consider a slab geometry of topological superconductors.
The normal part of the Hamiltonian can be written as
\bea
 H_0&=&
\sum_{\vk_{\perp},j}
\Big[c^{\dag}_{\vk_{\perp},j}h_{\perp}(\vk_{\perp})c^{}_{\vk_{\perp},j}
+c^{\dag}_{\vk_{\perp},j+1}h_zc^{}_{\vk_{\perp},j} \nnn
&&\qquad\qquad\qquad\qquad\qquad\ \ 
+c^{\dag}_{\vk_{\perp},j}h_z^{\dag}c^{}_{\vk_{\perp},j+1}\Big],
\label{hamiltonian_slab}
\eea
where $\vk_{\perp}=(k_x,k_y)$, $j$ is the position in $z$ direction: $j=L_z$ and $j=1$ correspond to the top and bottom surfaces, respectively.
As in eq. (\ref{hamiltonian-1}), we omit the band and spin indices.
In eq.(\ref{hamiltonian_slab}), 
\bea
 h_{\perp}(\vk_{\perp})&=&\sum_{a=1,2}R_a(\vk_{\perp})\alpha_a+M'(\vk_{\perp})\beta
+\epsilon'(\vk_{\perp})I 
\eea
and
\bea
 h_z&=& \frac{i}{2}A_3\alpha_3+B_1\beta-D_1I
\eea
describe spin-dependent hopping in $x$-$y$ plane and in $z$ direction, respectively, where
\bea
M'(\vk)&=&M_0-2B_1-\frac{4}{3}B_2 
\Big[3-\cos\big(k_ya\big) \nnn
&& \qquad\quad\quad
-2\cos\Big(\frac{\sqrt{3}}{2}k_xa\Big)\cos\Big(\frac{1}{2}k_ya\Big) \Big]\\
\epsilon'(\vk)&=&
-\mu+2D_1 
+\frac{4}{3}D_2
\Big[3-\cos\big(k_ya\big) \nnn
&& \qquad\quad\quad
-2\cos\Big(\frac{\sqrt{3}}{2}k_xa\Big)\cos\Big(\frac{1}{2}k_ya\Big)\Big].
\eea
The pair-potential part, $\hat\Delta$, 
is independent of both the momentum and the position.
We obtain the matrix element of the total Hamiltonian in the basis of $\{|{\vk_{\perp},j}\rangle\}$. 
By diagonalizing the Hamiltonian matrix, eigenenergies $E_{n,\vk_{\perp}}$ and eigenstates $|{n,\vk_{\perp}}\rangle$ are obtained.

\begin{figure}[h]
\begin{center}
\includegraphics[width=0.5\textwidth]{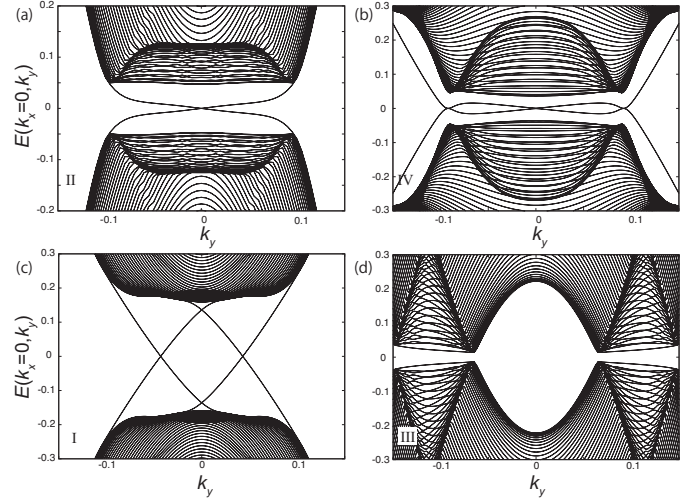}
\caption{
Energy spectrum of surface modes.
(a) $E_F=0.4$, $M_0=0.28$
(b) $E_F=0.5$, $M_0=1.0$
(c) $E_F=0.2$, $M_0=0.4$
(d) $E_F=0.9$, $M_0=0.28$.
The pair-potential in the bulk is fixed at $\Delta=0.05$ [eV].
The number of stuck layers 
in $z$ direction is $50$.
}
\label{Ek}
\end{center}
\end{figure}

Typical results of the spectrum are shown in Fig. \ref{Ek} for (a) phase II, (b) phase IV, (c) phase I, (d) phase III as labeled in Fig. \ref{mu-m0}.
The presence/absence of the surface modes indicates that those are topological/trivial superconducting states.
In Fig. \ref{Ek} (a) simple Majorana cone is seen as expected from previous work\cite{Fu2010}. 
As the chemical potential closes to the bottom of the conduction band, the surface spectrum deform from the simple Majorana cone as discussed in Refs. [\onlinecite{Hsieh2012}] and [\onlinecite{Yamakage2012}].
Such deformed surface Majorana mode can be seen also in phase IV (Fig. \ref{Ek} (b))\cite{Sasaki2011}. 
In phase I Majorana modes are doubled. This can be understood as follows.
In the normal limit, $\Delta=0$, phase I is a topological insulating state which has a single Dirac cone on the surface. As a pair-potential $\Delta$ is induced, the complex fermions are split into two independent Majorana modes.
A similar situation has been discussed in a quantum anomalous Hall insulator/ spin-singlet $s$-wave superconductor hybrid system\cite{Qi2010}.
In the next section, we discuss the connection between these surface modes and the quantized thermal Hall conductivity.


\section{Anomalous Thermal Hall conductivity}

\begin{figure}[t]
\begin{center}
\includegraphics[width=8cm]{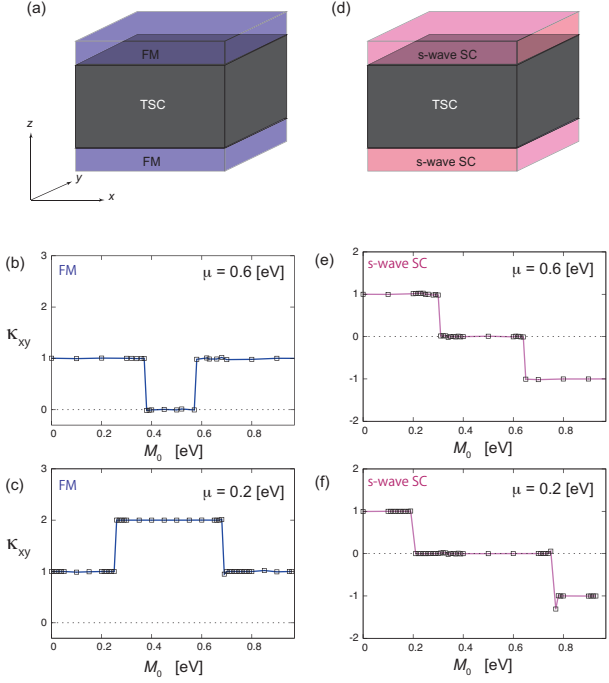}
\caption{(Color online)
Top: Illustrating the experimental setting for the measurement of the quantum thermal anomalous Hall effect of Majorana fermions on the surface of a three-dimensional topological superconductor with (a) attached ferromagnetic insulators and (d) attached $s$-wave superconductors with complex pair-potentials.
Bottom: Thermal Hall conductivity $\kappa_{xy}$ as a function of the band gap $M_0$ in units of $\pi^2k_B^2T/6h$ is plotted in (b), (c), (e), and (f).
The case of ferromagnets attached on the top and bottom surfaces is shown in (b) $\mu=0.6$ [eV] and (c) $\mu=0.2$ [eV]. The bulk pair-potential is fixed at $\Delta=0.05$ [eV].
$\kappa_{xy}$ with $s$-wave pairing induced by the proximity effect
is shown in (e) $\mu=0.6$ [eV] and (f) $\mu=0.2$ [eV].
}
\label{kappa}
\end{center}
\end{figure}

When the surface modes have a finite gap, the quantum thermal Hall effect occurs.
Physically the gap is induced by the magnetic interactions.
Figure \ref{kappa} (a) depicts an experimental setup where the interactions between quasiparticles and magnetizations in ferromagnets (or magnetic dopants such as Cr as in Ref.[\onlinecite{Exp_QAHE}]
) near the surface are introduced. A similar situation is realized by applying an external magnetic field. Due to the Meissner effect, the field can be finite within the penetration depth from the surface.
The magnetic interaction term is given by
\bea
 H_{\rm exc}=
\sum_{\vk_{\perp},j}
c^{\dag}_{\vk_{\perp},j}
J{M}_z(j)
\begin{pmatrix}
\sigma_z & 0 \\
0 & \sigma_z
\end{pmatrix}
c^{}_{\vk_{\perp},j},
\eea
where
$J$ is the exchange interaction constant and $M_z$ is the mean value of magnetic moments. 
For simplicity, we consider the case where magnetic moments are finite only at the top ($j=1$) and bottom ($j=L_z$) surfaces, while zero in the bulk.

To compute the thermal Hall conductivity, we use the generalized Wiedemann-Franz law\cite{Smrcka1977,Yokoyama2011} to the case of Majorana fermions\cite{note1,Sumiyoshi2013}; 
\bea
 \kappa_{xy}&=&\frac{\hbar\pi^2 k_B^2 T}{6L^2}
\sum_{n.m}\sum_{\vk_{\perp}}\theta(-E_{n,\vk_{\perp}}) \nnn
&& \quad
\frac{2{\rm Im}[\langle n,\vk_{\perp}|v_x|m,\vk_{\perp}\rangle \langle m,\vk_{\perp}|v_y|n,\vk_{\perp}\rangle]}{(E_{n,\vk_{\perp}}-E_{m,\vk_{\perp}})^2},
\label{WF}
\eea
where $L^2$ is the area of the surface.
Apart from the factor
$\pi^2 k^2_B T/6$,
the right hand side resembles the Kubo formula
for the electrical Hall conductivity,
which, however, is not a well-defined quantity for Majorana fermions;
nevertheless
Eq.\ (\ref{WF}) can be regarded as the generalized Wiedemann-Franz law
to Majorana fermions
\cite{Smrcka1977}.  
Compared to the electron systems, there is an extra factor of 
1/2 due to Majorana nature.

The thermal Hall conductivity $\kappa_{xy}$ is shown in Fig. \ref{kappa} as a function of the original band gap $M_0$.
Quantized values in units of $\frac{\pi^2k_B^2}{6h}T$ are clearly seen.
Figure \ref{kappa} (b) and (c) show $\kappa_{xy}$ in the presence of the magnetic interaction (Fig.\ref{kappa} (a)).
At the chemical potential $\mu=0.6$ [eV], the thermal Hall conductivity changes from $1\rightarrow 0\rightarrow 1$ as shown in Fig. \ref{kappa} (b) while the phase changes as II $\rightarrow$ III $\rightarrow$ IV.\cite{note2}
At $\mu=0.2$ [eV] (c), $\kappa_{xy}$ changes $1\rightarrow 2\rightarrow 1$, while the phase changes as II $\rightarrow$ I $\rightarrow$ IV. 
These results show that the quantized thermal Hall conductivity introduced by the magnetic interaction can distinguish phase I and III.
Moreover the quantized values characterize the number of surface Majorana dispersions.
The result $\kappa_{xy}=2\times \frac{\pi^2k_B^2}{6h}T$ in phase I is consistent to the fact that the surface modes in phase I are two split Majorana modes as shown in Fig. \ref{Ek} (c). Similarly, $\kappa_{xy}=0$ is consistent to the absence of surface modes in phase III, indicating that this phase is topologically trivial.

The other way to open a gap in the surface spectrum is introducing $s$-wave pairing with an imaginary pair-potential on the surface by the proximity effect with ordinary superconductors (SC) as proposed in Ref. [\onlinecite{Wang2011}] (Fig. \ref{kappa} (d)). 
In our model Hamiltonian, this interaction is described by 
\bea
 H_{s{\rm SC}}&=&
\sum_{\vk_{\perp},j}
\Big[c^{\dag}_{\vk_{\perp},j}
\Delta_s(j)
\begin{pmatrix}
\sigma_y & 0 \\
0 & \sigma_y
\end{pmatrix}
c^{\dag T}_{\vk_{\perp},j}
\nnn
&&\qquad +
c^{T}_{\vk_{\perp},j}
\Delta_s(j)
\begin{pmatrix}
\sigma_y & 0 \\
0 & \sigma_y
\end{pmatrix}
c^{}_{\vk_{\perp},j}\Big].
\eea
Because of the proximity effect, we assume that $\Delta_s$ is finite only at the top and bottom surfaces. When the $s$-wave pair-potentials, $\Delta_s$, have opposite sign between top and bottom, $\kappa_{xy}$ becomes finite.
Figure \ref{kappa} (e) and (f) both show that $\kappa_{xy}$ changes as $1\rightarrow 0\rightarrow -1$. 
At the chemical potential $\mu=0.6$ (e), the phase changes as II 
$\rightarrow$ III $\rightarrow$ IV. 
On the other hand, at the chemical potential $\mu=0.2$ (f), the phase changes as II $\rightarrow$ I $\rightarrow$ IV.
The results shown in Fig. \ref{kappa} (e) and (f) indicate that $\kappa_{xy}$ vanishes in phase I and III. The quantum thermal Hall effect induced by $s$-wave pair-potential does not distinguish the topological nature of phase I and III.

Let us compare the thermal Hall conductivity obtained above and the bulk topological number $\nu$.
As shown in Fig. \ref{nu}, $\nu$ changes as 1 $\to$ 0 $\to$ $-1$ as the phase changes II $\to$ III $\to$ IV and also II $\to$ I $\to$ IV. These behaviors are identified to the quantization rule of $\kappa_{xy}$ when the TSC is attached to $s$-wave SC with the complex pair-potential, consistent with Ref.[\onlinecite{Wang2011}]. 
On the other hand, the quantized value of the thermal Hall conductivity induced by the magnetic interaction 
should have some physical meanings. 
Recent theoretical work\cite{Fu2011,Ueno2013,Teo2013} addressed that an additional symmetry such as reflection has a role to protect gapless boundary modes characterized by the mirror Chern number, even in topologically trivial phases.
Instead of the bulk topological number $\nu$, the integers appeared in $\kappa_{xy}$ could be connected to the mirror Chern numbers. 

Quantum thermal Hall effect obtained above is understood in the viewpoint of surface Dirac theory, as follows.
We focus on Dirac fermions on the top surface.
\begin{figure*}
\centering
\includegraphics[scale=0.95]{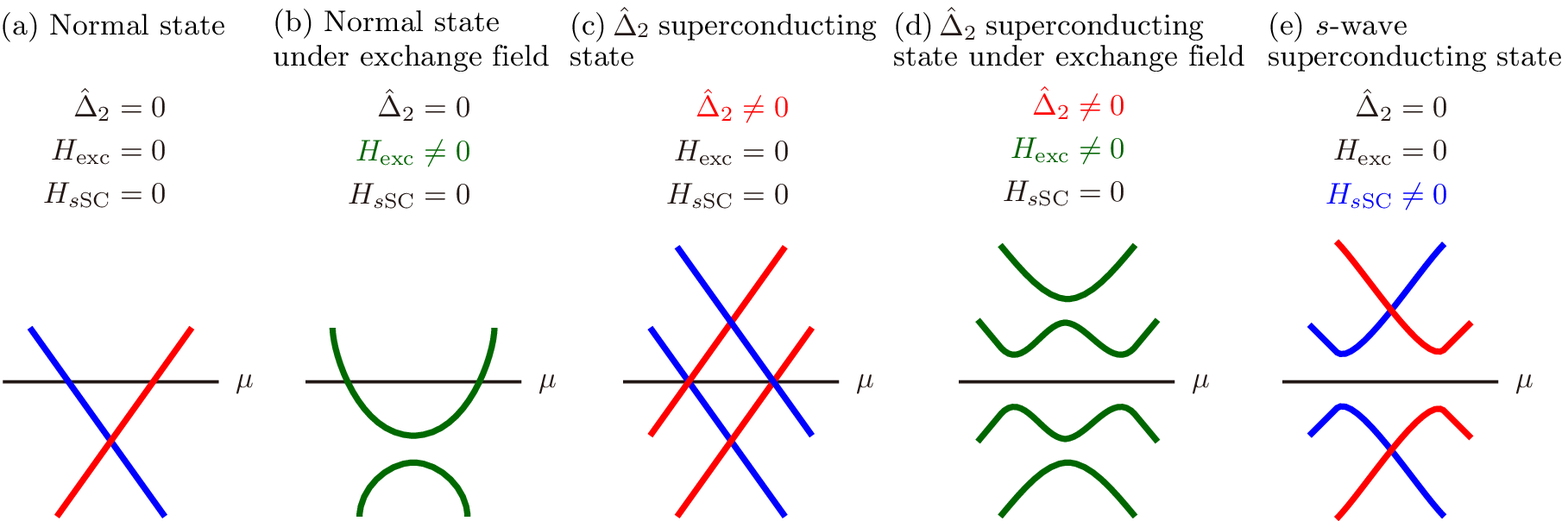}
\caption{Evolution of energy gap in the vicinity of the Fermi level $\mu$ from the $|N_{\rm M}|=2$ phase.}
\label{evogap}
\end{figure*}
Dirac fermions on the bottom surface gives basically the same results as those in the top surface.
The system in the normal state supports a surface Dirac fermion, as shown in Fig. \ref{evogap}(a), protected by time-reversal symmetry.
The surface effective Hamiltonian $H_{\rm surf}$ reads
\begin{align}
 H_{\rm surf}(\bm k) = v(k_x \sigma_y - k_y \sigma_x).
\end{align}
This Dirac fermion is also protected by mirror-reflection symmetry, and the corresponding topological number is given by the mirror Chern number $N_{M}^0$, which is defined by
\begin{align}
 N_{M}^{0} = \int \frac{dk_y dk_z}{2\pi} 
	\left[
	\bm\partial_{\bm k} \times 
	\bm A_{M}^{0}(k_y, k_z)
	\right]_x,
\end{align}
and
\begin{align}
&
 \bm A_{M}^{0}(k_y, k_z) 
\nonumber\\ &
= -\frac{i}{2} 
\sum_{n \in \mathrm{valence}}
\langle u_n^{0}(0, k_y, k_z)| 
	M \bm \partial_{\bm k} 
| u_n^{0}(0, k_y, k_z) \rangle.
\end{align}
Here, $|u_n^{0}(\bm k) \rangle$ is an eigenstate of the Hamiltonian $h(\bm k)$ for the normal state, the summation with respect to band indices $n$ runs over the valence bands, and $M$ is the mirror-reflection operator satisfying
\begin{align}
 M^\dag h(-k_x, k_y, k_z) M = h(k_x, k_y, k_z), 
 \
 M^2 = 1.
\end{align}
The topological insulating phase is characterized by $\left|N_{M}^{0}\right| = 1$.
The mirror Chern number $N_M$ for the superconducting states are defined by replacing $\bm A^{0}_M$ with
\begin{align}
&
 \bm A_M(k_y, k_z) 
\nonumber\\ &
= -\frac{i}{2} \sum_{n} \langle u_n^-(0, k_y, k_z) | \tilde M \bm \partial_{\bm k} | u_n^-(0, k_y, k_z) \rangle,
\end{align}
where the mirror-reflection operator $\tilde M$ in the $\hat \Delta_2$ superconducting state is given by
\begin{align}
 \tilde M = \begin{pmatrix}
  M & 0
  \\
  0 & -M^*
 \end{pmatrix},
\end{align}
and satisfies
\begin{align}
 \tilde M^\dag H_{\rm BdG}(-k_x, k_y, k_z) \tilde M = H_{\rm BdG}(k_x, k_y, k_z),
\end{align}
because of $M^{\rm T} \hat \Delta_2 M = -\hat \Delta_2$.\cite{Hsieh2012}
In the $\hat \Delta_2$ superconducting state, the mirror Chern number becomes twice $|N_{M}|=2$ as large as that in the normal state when the Fermi level is located in the bulk insulating band gap.\cite{Hsieh2012}
Consequently, two Dirac fermions show up on the surface of the superconductor [Fig. \ref{evogap}(c)], on the mirror-reflection symmetric line $k_x=0$.
The magnetic interaction $H_{\rm exc}$, which breaks the mirror-reflection symmetry, induces a mass gap in the Dirac fermions [Fig. \ref{evogap}(d)], 
and results in the quantum thermal Hall effect of $\kappa_{xy}^{\rm top} = |N_{M}| \kappa_0/2$ with $\kappa_0 = \pi^2 k_B^2 /(6h)$ for $|N_{M}|=2$. 
Summing up the contributions from top and bottom surfaces, one obtains $\kappa_{xy} = |N_{M}| \kappa_0$.
Imaginary $s$-wave pair potential $H_{s \mathrm{SC}}$ also opens a gap in the Dirac fermions.
This gapped surface state, however, is in a topologically trivial phase, in the sense that as one turns off the topological superconducting gap $\hat \Delta_2$,  
the state is continuously connected to a surface $s$-wave superconducting state  ($H_{s \mathrm{SC}} \ne 0$ and $\hat \Delta_2 = 0$)
without gap closing owing to $\{ H_{s \mathrm{SC}}, \hat \Delta_2 \} = 0$. 
The surface $s$-wave superconducting state is described by the Hamiltonian $H_{s \mathrm{surf}}$:
\begin{align}
 H_{s\mathrm{-surf}}(\bm k) 
 =
 \begin{pmatrix}
	H_{\rm surf}(\bm k)-\mu &  \sigma_y \Delta_s
	\\
	\sigma_y \Delta_s & -H^{\rm T}_{\rm surf}(-\bm k) + \mu 
 \end{pmatrix},
\end{align}
and the corresponding surface energy spectrum is illustrated in Fig. \ref{evogap}(e). 
Therefore, one can conclude that the thermal Hall conductance vanishes in the $|N_{M}|=2$ phase with an imaginary $s$-wave pair potential.
Note that the above discussion is valid only for the $|N_{M}|=2$ phase, where the Fermi level is located within the topological insulating gap.
In the $|N_{M}|=1$ and $|\nu|=1$ phase, 
since the normal state $\hat \Delta_2=0$ is gapless, i.e., bulk metallic, the surface Dirac theory in the normal state no longer stands and
the bulk electronic states have to be taken into account.
Eventually, the massive Majorana surface states induced by $H_{s \mathrm{SC}}$ in the $|N_{M}|=1$ and $|\nu|=1$ phase are not continuously deformed into trivial gapped surface states.
As a result, 
the system is in a nontrivial phase and gives $\kappa_{xy}^{\rm top} = \nu\kappa_0/2$, and $\kappa_{xy} = \nu \kappa_0$ in the whole system.

Again, let us mention that the magnetically coupled system shows the quantum thermal Hall effect.
In the normal state, the magnetic interaction opens a gap at the surface Dirac point below the Fermi level, as shown in Fig. \ref{evogap}(b).
Thus, the surface states for $H_{\rm exc} \ne 0$ and $\hat \Delta_2 = 0$ is gapless at the Fermi level.
This means that the gapped surface states induced by $H_{\rm exc} $ with the help of $\hat \Delta_2$ is not connected to a trivial one 
as $\hat \Delta_2$ is switched off.
Therefore, the system is in the topologically nontrivial phase that exhibits the quantum thermal Hall effect of $\kappa_{xy} = |N_{M}| \kappa_0$.


\section{Discussion}

Time-reversal invariant topological superconductors in three-dimensions (class DIII) are characterized by integers ${\mathbb Z}$ in contrast to topological insulators (class AII) which have ${\mathbb Z}_2$ numbers.\cite{Schnyder2008,Kitaev2009}. In this paper we studied the relation between the topological invariant and the thermal Hall conductivity, basing on the model for superconducting states in Cu-doped Bi$_2$Se$_3$. The thermal Hall conductivity shows different behavior when the surface gap is induced by the magnetic interactions and by the complex $s$-wave pair-potentials.
As discussed in Ref. [\onlinecite{Wang2011}], we confirmed that the topological invariants of time-reversal invariant superconductors in three dimensions are directly related to the surface thermal Hall conductance induced by the complex $s$-wave pair-potentials.

The quantum thermal Hall effect on the surface of topological superconductors are closely related to the cross-correlated thermal responses of topological superconductors. Since gauge symmetry is spontaneously broken in superconducting phases, it is useful to consider responses to the gravitational fields instead of electromagnetic fields.
 In Refs. [\onlinecite{Ryu2011}] and [\onlinecite{Wang2011}] the gravitational instanton term was introduced to characterize the thermal responses, because the gravitational fields can be connected to the thermal gradients\cite{Luttinger1964}. It was argued that the gravitational instanton term can characterize only the ${\mathbb Z}_2$ part of topological classification\cite{Wang2011}.
Later the energy density functional which characterizes the cross-correlated responses between the thermal gradient and mechanical rotational motion with full ${\mathbb Z}$ classification of three-dimensional topological superconductors was derived in Ref.[\onlinecite{Nomura2012}]. The strength of these couplings, however, were found to be very small, and thus the effect may not be measurable.
The present work concludes that topological invariants of three-dimensional topological superconductors can be measured by thermal transport on the surface subjected to the proximity with complex $s$-wave pairing field.

\section*{Acknowledgments}

The authors gratefully acknowledge A. Furusaki, T. Ito, D. Kurebayashi, R. Nakai, S. Ryu, M. Sato, and K. Shiozaki for valuable comments and stimulating discussions.
This work was
supported by
the ``Topological Quantum Phenomena" (No. 25103703) Grant-in Aid for Scientific Research
on Innovative Areas 
and by
Grant-in-Aid for Scientific Research (No. 24740211)
from the Ministry of Education,
Culture, Sports, Science and Technology (MEXT)
of Japan.



\begin{thebibliography}{10}
\bibitem{TSF} 
We do not distinguish TSFs from TSCs henceforth,
because thermal responses are generic
and common to both of them.
\bibitem{note2}
When the width of the system is not large enough, 
the positions where $\kappa_{xy}$ changes differ from the phase transition points of the bulk system.
\end{thebibliography}
\end{document}